\documentclass{article}

\usepackage{arxiv}

\usepackage[utf8]{inputenc} 
\usepackage[T1]{fontenc}    
\usepackage{hyperref}       
\usepackage{url}            
\usepackage{booktabs}       
\usepackage{amsfonts}       
\usepackage{nicefrac}       
\usepackage{microtype}      
\usepackage{lipsum}
\usepackage{graphicx}

\title{Diffusion-driven demographics – Turing model as a concept for the emergence of sedentism}

\author{
  John Friesen, Jakob Hartig, Katharina Henn and Peter F.~Pelz \\
  Chair Fluid Systems\\
  Technical University of Darmstadt\\
  Darmstadt, 64285 \\
  \texttt{john.friesen@fst.tu-darmstadt.de} \\
}

\begin{document}
\maketitle

\begin{abstract}
Sedentism was a decisive moment in the history of humankind. In a review article Kay and Kaplan quantified land use for early human settlements and found that sedentism and the emergence of farming go hand in hand. For these settlements two primary land use categories, farming and living, can be identified, whereas for hunter gatherer societies no distinct differences can be made. It is natural to search for this in the behavior of two different groups, settlers and farmers. The development of two distinct zones and the two groups lead us to the hypothesis that the emergence of settlements is the result of diffusion-driven Turing instability. In this short communication we further specify this and show that this results in a regular settlement arrangement as can still be seen today in agricultural regions.
\end{abstract}


\section{Introduction}
Over the last ten millennia, humanity has changed significantly. While 10,000 years ago we were primarily hunters and gatherers, the last 5,000 years have seen the emergence of smaller settlements that have since grown steadily and now house more than half of our species \cite{taylor_extraordinary_2012}. The sedentism of people has been modelled qualitatively in numerous publications e.g. \cite{kelly_mobilitysedentism_1992}. By formulating the models verbally, counterintuitive behaviours and properties due to the complex nature of the process may be omitted. Consequently, quantifiable models are a useful alternative for understanding complex human processes \cite{turchin_war_2013}. 
There are repeated attempts to quantify history as well as the political, economic and social processes shaping it, like Richardson \cite{richardson_mathematical_1935} describing the psychology of war with formulas or Krugman \cite{krugman_self-organizing_1996}, who used a Turing model to describe the spatial distribution of industry and economy. In addition to these attempts to describe human behaviour quantitatively, there are approaches in which human sciences use scientific phenomena to explain sociological processes, like Luhmann's systems theory \cite{luhmann_soziale_2018}, which uses the concept of autopoiesis from biology \cite{varela_autopoiesis_1974}. The combination of human sciences and sciences thus holds great potential.

In this paper, we present a model for sedentism based on the well-known model of morphogenesis by Alan Turing \cite{turing_chemical_1952}. Turing thereby answered the question how patterns emerge in originally homogeneous biological systems. More precisely, he stated how the homogeneous distribution of cells shortly after fertilization of the ovum develops into structures, e.g. extremities, based on an instability caused by diffusion. 
Despite the simplifications and assumptions made in the modelling process, quantitative mathematical models are very useful, as Kondo and Miura \cite{kondo_reaction-diffusion_2010} put it eloquently: \textit{“The logic of pattern formation can be understood with simple models, and by adapting this logic to complex biological phenomena, it becomes easier to extract the essence of the underlying mechanisms.”}

\section{Model}
\begin{figure}[h]
  \centering
  \includegraphics[width=\textwidth]{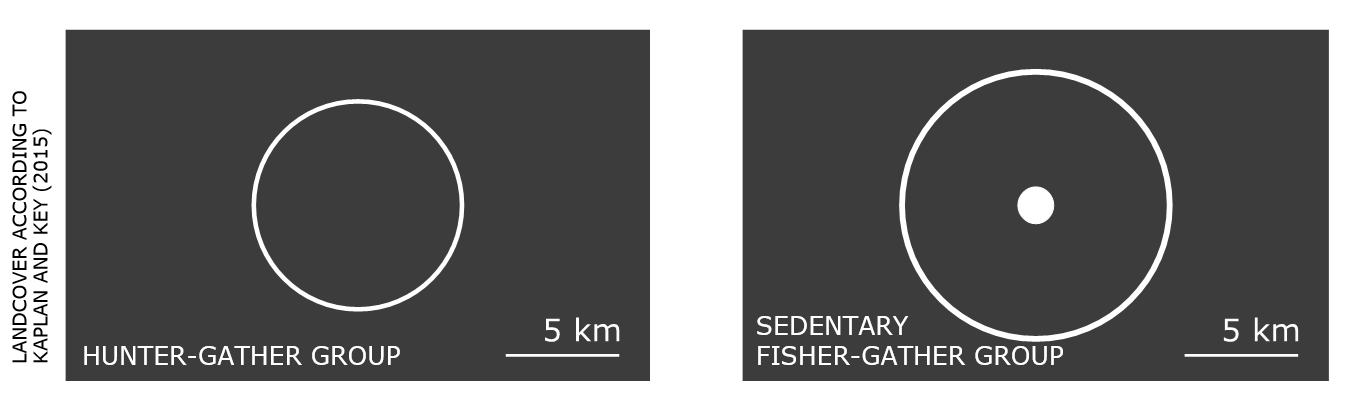}
  \caption{Two schematic representations of land use at different times of settlement, adapted from \cite{kay_human_2015}. The grey area represents fertile land, the white circle on the right represents a settlement and the circles in both figures represent the area where the respective communities hunted or farmed.}
  \label{fig:fig1}
\end{figure}

In order to grasp the essence of sedentism we transfer Turing’s idea to homogeneous landscapes, describing geographical regions with low altitude and fertile soil \cite{yang_spatial_2016}, and ask how settlements developed. To this end, we first look at the process of sedentism from an archaeological perspective: Kay and Kaplan \cite{kay_human_2015} classified and quantified land use and its development of human societies in Africa in the period between 1000 BC and 1500 AD at different points in time. According to their study, sedentism took place in a development from hunter-gatherer communities that did not establish permanent settlements in their habitat (Fig. \ref{fig:fig1}, left) to farmers societies that significantly transformed part of their habitat by establishing permanent settlements (Fig. \ref{fig:fig1}, right), so that remains of these settlements can still be identified by excavations today.

Our hypothesis is that settlements in homogeneous rural areas (e.g. regions like the Nile delta) are the result of a Turing mechanism occurring through the interaction of two groups. This hypothesis is not so far-fetched as it might seem at first glance since Turing patterns do not just emerge in chemical systems \cite{vigil_turing_1992} or in patterning of organisms \cite{murray_mathematical_2003}. Theraulaz et al. \cite{theraulaz_spatial_2002} show that spatial clustering in ant colonies is a Turing pattern, emerging through self-organization of these social insects. This shows that Turing patterns do not solely emerge in organisms but also through the interaction of organisms.
Patterns in the animal world are in some cases seen as indirect Turing patterns, the visible patterning is no Turing pattern itself. The patterns are believed to be laid down in the stadium of an embryo by the interaction of so-called morphogens. This prepattern is then frozen by different chemical reactions and becomes visible with the development of the animals features \cite{murray_mathematical_2003}. 
In the presented scheme of land use in Fig. 1 there are two dominant anthropogenic structures: agricultural land and settlements. It seems obvious to search for the reason for this two land use types within the interaction of two somehow different groups of people. The two groups are called “settlers” and “farmers” in the following, which implies a form of specialization of profession in society, because division of labour is seen as one of the preconditions and consequences of sedentism \cite{haaland_sedentism_1995}. Just like in the world of animals, the observed patterns of two dominant anthropogenic structures in rural settlements are not directly the Turing-patterns but the result of a Turing pattern of the morphogens “settlers” and “farmers”. 
In contrast to agent-based modelling, a well-known approach, the Turing-mechanism directly is a model for the sum of actions of many actors. We relate the settlements with the settlers and the fields with the farmers by considering the probability of presence $c_i$ of the settlers ($i=1$) and farmers ($i=2$). The interaction of these two groups can be summarized in the form of two coupled equations for the probabilities of presence $c_i$:
\begin{equation}
\frac{\partial c_i}{\partial t}=f_i(c_j).
\end{equation}

To grasp the complex social interactions in a model, there have to be simplifications. Therefore, we identified a simple set of rules and conditions that we believe are central to the development of settlements in homogeneous fertile land \cite{pelz_similar_2019}. At the beginning of sedentism there is a homogeneous probability of presence for settlers and farmers and the primary economic process is self-sufficency by hunting and gathering. The rules are:
\begin{itemize}
\item	Humans are social beings and search for communities to create synergies. Consequently, the change in the probability of settlers' presence increases with the probability of settlers' presence.  
\item The more food is available, the more attractive the place is for settling. Thus, the change in the probability of settlers' presence increases with the probability of farmers' presence. 
\item The more settlers are at one place, the less attractive this place is for farmers, since it is impossible to farm within regions with a high probability of settlers’ presence. Therefore, the change in the probability of farmers' presence decreases with the probability of settlers' presence.  
\item The yield per surface area of arable land is limited and each farmer strives to maximize his yield. Consequently, the change in the probability of farmers' presence decreases with the probability of farmers' presence. It should be explicitly mentioned here that the farmers also have an interest in communities and live in settlements, but due to their work in the fields they do not have a high probability of staying in the settlements.
\end{itemize}
The latter presented conditions can be expressed in a mathematical way and represent a stable state, when Eq. 1 satisfies the following conditions

\begin{equation}
\left(\frac{\partial f_i}{\partial c_j}\right)_{2,2}=\left(a_{i,j}\right)_{2,2}=\left(\begin{array}{cc} a_{11}>0 & a_{12}>0\\ a_{21}<0 & a_{22}<0 \end{array}\right).
\end{equation}

Until now, the spatial movement of the two groups is not considered. We include this by assuming that both groups have mobilities $\mu_i$ which can be used to derive diffusion coefficients $D_i=\mu_i k_B T$ using Einstein’s relation \cite{einstein_motion_1905} as an analogy leading to  

\begin{equation}
\frac{\partial c_i}{\partial t}=f_i(c_j)+d_{i,j}\frac{\partial^2 c_j}{\partial x_k \partial x_k},(d_{i,j})=\left(\begin{array}{cc} a_{11}>0 & a_{12}>0\\ a_{21}<0 & a_{22}<0 \end{array}\right).
\end{equation}

$d$ is the ratio of diffusion coefficients, i.e. the ratio of mobilities $D_2/D_1=\mu_2/\mu_1$. Due to the low division of labour in the Palaeolithic, it can be assumed here that both groups exhibit similar mobility ($d=1$), leading to a steady-state with no pattern formation.

In the next step, we assume that the mobility of settlers and farmers begins to differ from each other. At first it is irrelevant whether the diffusion coefficient and thus the mobility of the farmers increases and they can cultivate more land in a certain time, or the mobility of the settlers decreases, since they become more "immobile" due to the division of labour. The system is in a stable steady-state, as long as the diffusion coefficients stay similar $d<d_{crit}$. 

\begin{figure}[t]
  \centering
  \includegraphics[width=\textwidth]{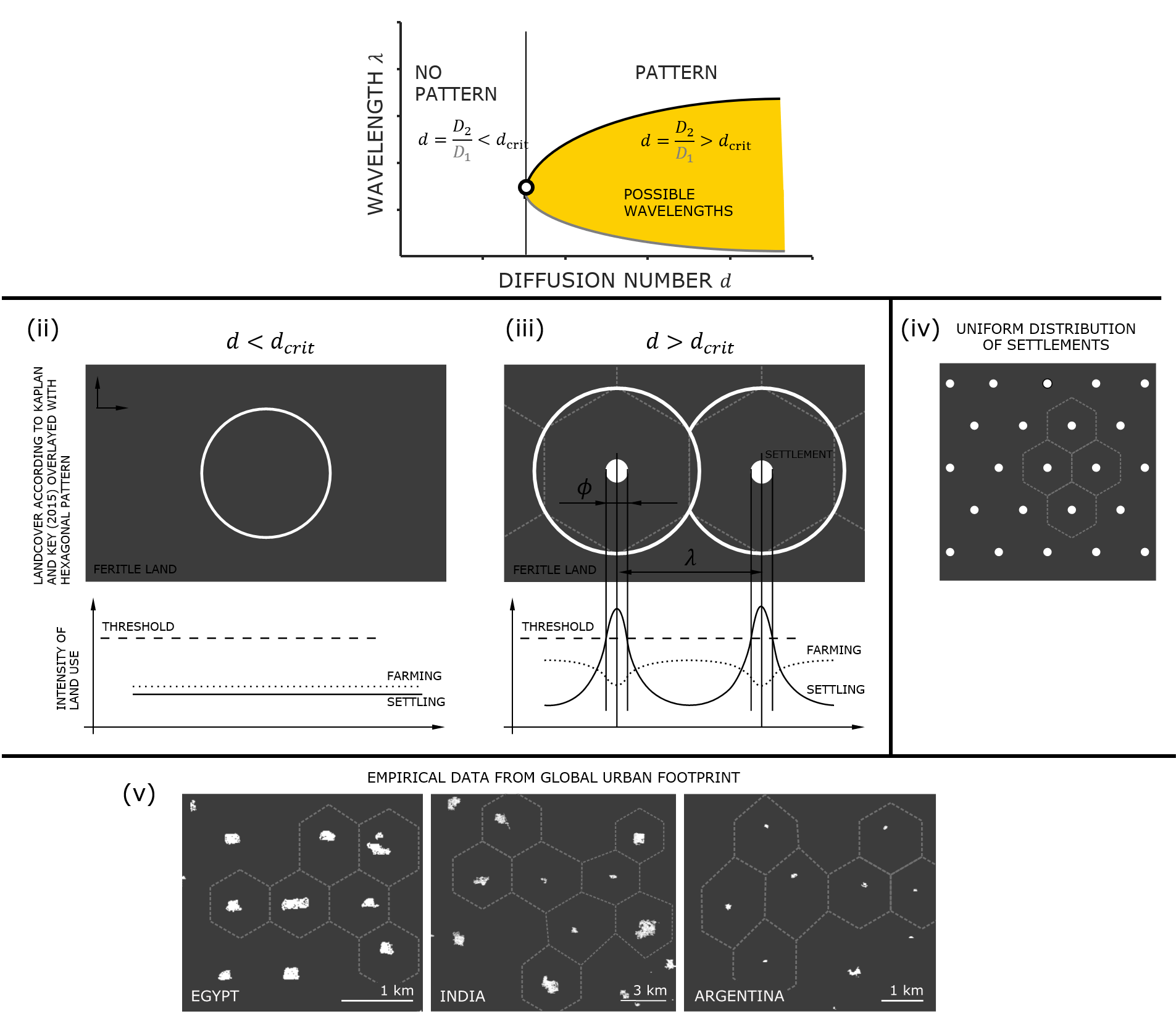}
  \caption{Bifurcation due to different diffusion coefficients (i) and land cover (adapted from \cite{kay_human_2015}) and the connected intensity of land use for hunter-gatherer (ii) and urbanized societies (iii). Theoretical uniform settlement distribution (iv) and empirical settlement patterns (v) in Egypt, India and Argentina derived from the Global Urban Footprint \cite{esch_where_2018, esch_breaking_2017}}.
  \label{fig:fig2}
\end{figure}

However, if the diffusion coefficient exceeds a threshold value (for $d>d_{crit}$), the system behaviour changes fundamentally, as the system migrates into a bifurcation, cf. Fig. \ref{fig:fig2} (i). The previously stabilizing effects are cancelled out by the changed mobility of the morphogens. As Turing impressively showed, the diffusion process, which usually leads to a uniform distribution of morphogens, leads under certain conditions to instability of the morphogenic interactions and thus to the formation of concentration peaks.

We further assume that exceeding the probability of settlers to stay above a threshold value results in the establishment of settlements of the size $\phi$. Again, we point out the above-mentioned analogy to the formation of animal fur patterns. According to our thesis, the settlement structures are not Turing patterns themselves, but only the result of the strong concentration peaks generated by the Turing mechanism.
These results confirmed the findings of Kay and Kaplan \cite{kay_human_2015} presented in Fig. \ref{fig:fig1}. Hunter-gatherer communities use a defined agricultural area (circle in Fig. \ref{fig:fig2} (ii)) but do not establish permanent settlements, whereas urbanized societies also use a defined area for agriculture but establish permanent settlements (white and filled white circle respectively in Fig. \ref{fig:fig2} (iii)).

Hence, the emergence of settlement structures may be the result of a diffusion-driven instability. 
However, the hypothesis presented here is not only an approach to explain the formation of settlement structures, but also their spatial distribution on a homogeneous landscape. The mobility of the two groups described above, as well as their interaction, leads to a regular distribution of settlements, with a distance $\lambda$ to each other (Fig. \ref{fig:fig2}, (iii)). The radius determined by Kay and Kaplan is approximately half the wavelength of the model presented here. For the relatively simple case of river deltas, where geographical homogeneity is given, the resulting patterns will be regular hexagonal \cite{ouyang_transition_1991}.

This result is in line with empirical observations of today's agricultural areas, using satellite data of rural areas near rivers, where regular hexagonal settlement arrangements can be observed (cf. Fig. \ref{fig:fig2},(v) or \cite{yang_spatial_2016}). Although these settlements are most likely much younger than the settlement processes mentioned above, we put forward the hypothesis that older settlements were also arranged in this way. 
Also in geographical theories, such as Christaller's central place theory \cite{christaller_zentralen_1933}, the initial state of urbanization is assumed to be an uniform hexagonal distribution of settlements on an unbounded isotropic homogeneous space with an evenly distributed population (Fig. \ref{fig:fig2},(iv)).

\section{Discussion}

In this paper we presented a model that describes the sedentism of humans as a result of diffusion-driven instability, which insists on a changed mobility of different social groups. So far, the approach is to be understood purely qualitatively. In real settlement formations the surrounding topography naturally has an influence on settlement formation. Furthermore, detailed quantitative analyses are necessary to determine the reaction kinetics $f_j$ as well as the exact parameters for the mobilities or for d.
As already discussed by Kondo and Miura \cite{kondo_reaction-diffusion_2010}, the question arises to what extent the patterns discussed here are results of a Turing mechanism. In biological systems, the dynamics can often be investigated by intervening in the system and the behaviour to be observed afterwards. This is usually not possible in the case of settlement structures, since the processes assumedly have more complex dimensions considering the influence of different economic, ecological, political and social factors. But as stated in the beginning, with this quantifiable model the essence of the underlying processes can be grasped despite simplifications. Therefore, the possibility of a Turing mechanism being the reason for sedentism and regular settlement patterns should be considered in future research to gain further insight into sedentism of mankind.

\bibliographystyle{unsrt}  
\bibliography{references}  

\end{document}